\documentclass[fleqn,10pt]{wlscirep}
\usepackage{caption}
\usepackage{textcomp}
\usepackage[utf8]{inputenc}
\usepackage{newunicodechar}
\usepackage{chngcntr}
\newunicodechar{−}{-}
\usepackage{float}
\usepackage{array}
\usepackage{geometry}
\usepackage{multirow}
\usepackage[T1]{fontenc}
\usepackage{tikz}
\usepackage{todonotes}
\usepackage{textcomp}
\usepackage{tcolorbox}
\usepackage{textgreek}
\usepackage{graphicx}
\usepackage{subcaption}
\usepackage{xcolor}
\usepackage{underscore}

\newif\ifediting
\editingfalse

\usepackage{soul}

\ifediting
    
\else
    
    \renewcommand{\st}[1]{}
\fi

\title{Does online sustainability communication shape public discourse? Insights from six years of tenant-housing provider interactions}

\author[1,*]{Shray Juneja}
\author[2]{Suzan Verberne}
\author[1]{Ioulia V. Ossokina}
\affil[1]{Eindhoven University of Technology, the Netherlands}
\affil[2]{Leiden University, the Netherlands}

\affil[*]{s.juneja@tue.nl}

\begin{abstract}
Authorities increasingly rely on social media to advance sustainability transitions, infrastructure investment, and service reform. Yet how citizens respond to these digital communications remains poorly understood. Existing approaches rely on aggregate engagement metrics (e.g., likes), providing limited insight into discourse structure and quality. We developed a data-driven, multidimensional framework to analyse how social media communication shapes the content of discourse, focusing on sustainability-related engagement in Dutch public housing. We analysed 792 posts and 3,197 tenant comments from the Facebook pages of 92 housing providers (2018-2023). A machine-learning pipeline classified comments into recurring discourse configurations across three dimensions - communicative intent, sentiment, and semantic relatedness. Multinomial logistic regression estimated the effects of post-design and organisational characteristics on discourse. Tenant comments were significantly more semantically aligned with their corresponding posts than with randomly paired content, indicating that organisational communication structures responses to topics. Six discourse types emerged, with critical and inquiry-driven engagement increasing over time. Post-level features did not significantly explain variation; organisational characteristics dominated. Larger housing associations attracted more substantive responses, while lower-rent organisations received fewer evaluative comments. While applied to housing associations, our methodology provides a scalable approach to analyse online discourse dynamics, quality, and content across organisations and contexts.

\end{abstract}

\begin{document}

\flushbottom
\maketitle


\thispagestyle{empty}

\section{Introduction}
Social media has become a central arena through which organisations communicate with citizens on consequential public issues, including sustainability transitions, infrastructure investment, and service reform. In these contexts, communication is not merely informational but constitutive: it shapes how citizens interpret policy, evaluate legitimacy, and decide whether to engage \cite{bartling2024public,bail2018exposure}. A growing body of research shows that the content and quality of online communication influence both citizen participation \cite{zollo2026examining, nisar2018trains} and the effectiveness of organisational initiatives \cite{nguyen2024citizen, liu2023mapping}.

Most existing research has measured audience response using aggregate metrics such as likes, comments, and shares, which capture participation volume but obscure its qualitative nature \cite{leppert2022role, stone2020linguistic, Han2024government, liu2023mapping, zhang2022citizen, teneboim2022comments, lee2018advertising}. Under these metrics, distinct forms of engagement, such as critical evaluation, information seeking, or passive redistribution, can appear equivalent, even though responses vary in affect, communicative function, and alignment with the original message. The multidimensionality of organisational communication matters for legitimacy, trust, and policy outcomes \cite{grimmelikhuijsen2013effect, Contri_2025_Using}, yet existing approaches have failed to capture it, thereby limiting understanding of how organisational communication shapes public discourse.

Typology-based approaches present a promising alternative by classifying discourse into distinct modes of engagement. However, existing typologies have typically been qualitative or theoretically derived and have relied on corpus-based approaches to capture discourse variation \cite{biber2021towards, romero2021corpus, wukich2022social} or operationalisation along a single dimension, such as sentiment \cite{nisar2018trains} or topic modelling \cite{laureate2023systematic}. Consequently, these methods have captured only partial aspects of meaning while overlooking how multiple dimensions interact. There has been no scalable method that simultaneously captures citizens' affective orientation, communicative intent, and the degree to which their responses are substantively related to the issue.

To address this gap, we developed a data-driven framework that characterises discourse as a multidimensional construct and identifies empirically grounded discourse types. Specifically, we operationalised three dimensions, communicative intent, sentiment, and semantic relatedness, and used their joint distribution to identify recurring configurations of discourse types. By doing so, we captured aspects of discourse that were not observable through conventional engagement metrics and provided a more fine-grained account of how citizens respond to organisational communication. We further examined, using statistical analysis, how organisational characteristics and communication design influence the type of discourse that emerges. 

We applied this framework to sustainability communication by Dutch public housing associations, which are non-profit organisations owning and managing approximately 30\% of the national housing stock and providing rental housing to lower-income households. This setting provides an interesting empirical context: housing associations are legally required to secure approval from at least 70\% of affected tenants before implementing major (sustainability) retrofits \cite{ossokina2021does}. Communication thus functions as a governance mechanism, where citizen responses can directly enable or constrain policy implementation \cite{voorberg2015systematic, torfing2012}. Our dataset comprises 792 posts and 3,197 comments from 92 housing associations between 2018 and 2023.

This study makes two contributions. First, it introduces a scalable framework that jointly models semantic relatedness, sentiment, and communicative intent in social media comments, enabling the identification of discourse configurations not captured by conventional engagement metrics. Second, it analyses how the variation in discourse type is affected by organisational characteristics and communication design. While developed in the context of sustainability communication in public housing, the framework is transferable to a wide range of organisational and policy settings.

\section{Analytical Framework}
\label{sec:framework}

\subsection{Discourse typology}
Earlier literature showed that engagement and deliberation are inherently multidimensional, varying systematically across topical, behavioural, and affective dimensions \cite{Segev2023, Gibson2013}. The goal of our discourse typology was to meaningfully aggregate these multiple dimensions into a limited number of discourse configurations. We operationalised the three dimensions in a way that aligns with the policy information needs of local authorities and non-profit organisations such as housing associations \cite{simonofski2021supporting,condie2025tenant}, with sentiment, communicative intent, and semantic relatedness measuring the affective, behavioural, and topical dimensions respectively. Each comment was then represented within this space, allowing discourse types to be identified as recurring configurations of these dimensions.

Semantic relatedness captures the extent to which a comment engages with the content of the original post and was operationalised as cosine similarity between post and comment embeddings \cite{reimers-2019-sentence-bert, li2022impact}. Sentiment captures evaluative tone, which was measured using standard sentiment analysis \cite{liu2012}. Communicative intent captures functional orientation. Building on speech act theory \cite{austin1962how,searle1969speech}, we distinguished criticism, appreciation, inquiry, statement, and forwarding behaviour. We fine-tuned transformer models to classify comments into these classes.

\subsection{Organizational and communication drivers of discourse variation}
\label{org_comm}
In the next step, we assessed how communication design and organisational characteristics jointly shape the distribution of discourse configurations. A body of research shows that social media content design influences audience response. Structural features such as message length, lexical richness, and dialogic features (e.g., question posing) are expected to be associated with more substantive and interactive discourse \cite{zollo2026examining, wang2025too, hussin2025engaging}. Informational features (e.g., URL inclusion) are expected to promote more passive forms of engagement \cite{sabate2014factors}. Message framing also affects response: positive content tends to elicit supportive reactions, whereas negative or problem-focused messages more often trigger critical or emotionally charged responses \cite{lee2018advertising}. We operationalised communication features using message length, inclusion of questions, and URLs. 

Alongside post-level design, organisational characteristics may shape how audiences interpret and respond to communication, given evidence that structural context and online environments amplify evaluative responses to institutional actors \cite{wallace2021big, lovejoy2012information, mergel2013social}. For the case of housing associations, we included the following organisational features: scale (size of the housing stock owned), affordability of the rents, and tenant satisfaction.

Both classes of predictors were examined simultaneously in a combined multinomial logistic regression model. Overall, the framework provides a parsimonious and testable account of citizen discourse as a structured, multidimensional phenomenon shaped by communication design and organisational context.

\section{Data Collection \& Pre-processing}
We applied this framework to sustainability communication by Dutch public housing associations, non-profit organisations owning and managing around 30\% of the national housing stock (2.3 million homes in 2025) and providing rental housing to lower-income households. This setting offers a valuable empirical context, as housing associations are legally required to obtain approval from at least 70\% of affected tenants before implementing major sustainability retrofits.

We collected data from the public Facebook pages of 92 housing associations. Our dataset consists of posts published by housing associations and their associated residents' comments. The data spans from 03 January 2018 to 06 May 2023. We retained only posts that (i) had at least one comment and (ii) were related to the topic of sustainability. The latter was defined using a rule-based filter on Dutch terms for energy, insulation, heat, solar, renovation, water pump, and sustainability (strings \texttt{energ}, \texttt{isolat}, \texttt{warmt}, \texttt{zonne}, \texttt{duurz}, \texttt{renov}).

The Ethics Boards of Leiden University and Eindhoven University of Technology approved our study design and data management plan. In accordance with the GDPR (Article 9.2), individual consent was not obtained, as the GDPR permits the use of data from publicly accessible forums with justified cause. To protect user privacy, usernames were not collected, collected messages were stored securely, and access was restricted to involved researchers and annotators. For data labelling, we did not use commercial tools but set up private servers accessible only to annotators.

The data was preprocessed by removing replies/comments from housing associations, as our focus was on residents' voices. Empty rows or deleted posts were also removed. To protect user privacy, all occurrences of personal names and housing association names in the comments were anonymised by replacing them with the tokens \texttt{<PERSON>} and \texttt{<ORG>}, respectively. Named entity recognition (NER) was performed using the spaCy library\footnote{The spaCy NER model \texttt{nl_core_news_lg} was used for Dutch-language named entity recognition} for personal names, while housing association names were replaced using the list of 92 organisations compiled from the post authors present in the dataset. The resulting dataset consists of 792 posts and 3,197 comments. On average, each post received 4 comments, ranging from 1 to 50.

\section{Methods}
\subsection{Discourse Typology: Relatedness, Intent, Sentiment and Clustering}
To construct the discourse typology, we characterised each comment along three dimensions, semantic relatedness, sentiment, and communicative intent, and derived discourse types through unsupervised clustering of these dimensions. Each classification pipeline is described below, followed by the clustering procedure.

\subsubsection{Semantic Relatedness}
We operationalised semantic relatedness as the cosine similarity between post and comment sentence embeddings, yielding a continuous relatedness score in [−1, 1] for every post-comment pair.

\begin{figure}[t]
\centering
\includegraphics[scale=0.25]{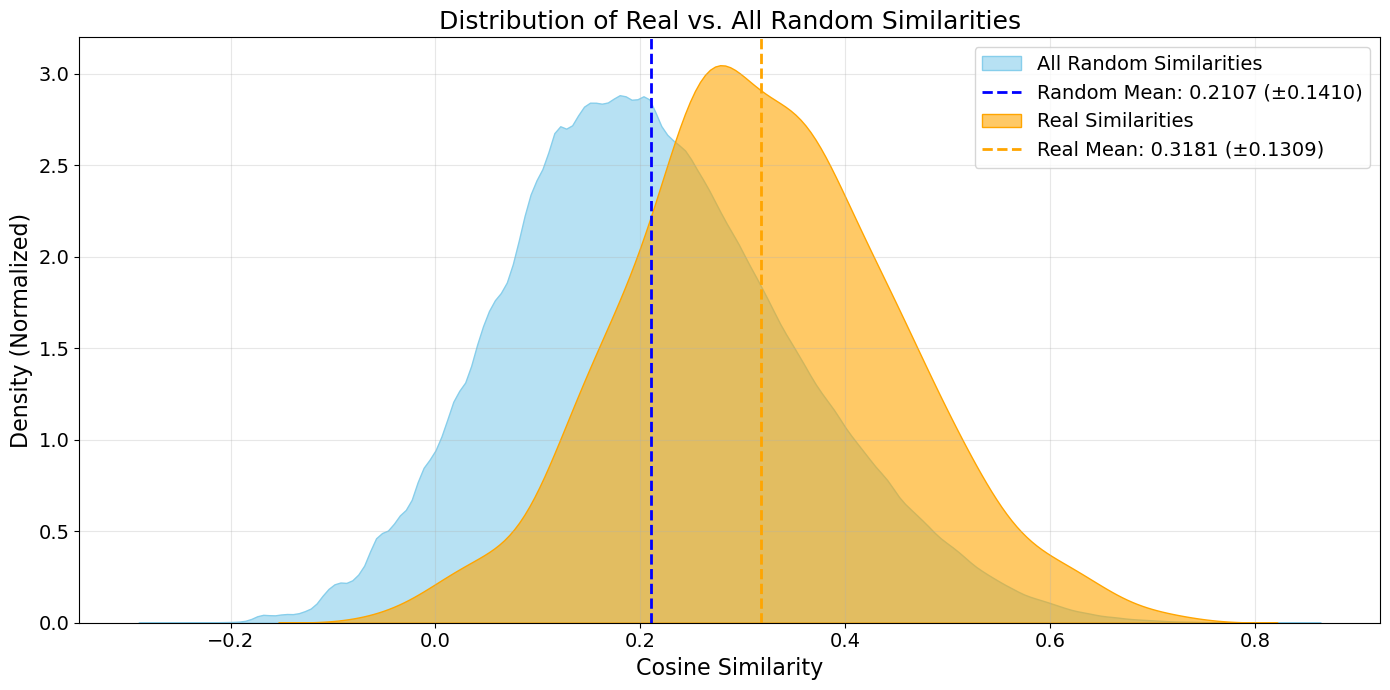}
\caption{Comparison of real post-comment similarity scores with similarities obtained from randomly paired comments.}
\label{fig:comments_postsl}
\end{figure}

Embeddings were generated using the \textit{paraphrase-multilingual-MiniLM-L12-v2} model,\footnote{\url{https://huggingface.co/sentence-transformers/paraphrase-multilingual-MiniLM-L12-v2}} a sentence transformer that maps variable-length text to fixed-dimensional dense vectors \cite{reimers-2019-sentence-bert}. We selected this model because it supports multilingual input, including Dutch, is optimised for semantic similarity tasks, and remains computationally efficient at scale. For each post and comment, we independently generated a 384-dimensional sentence embedding and computed the cosine similarity between the post embedding and its associated comment embedding.
To verify that comments were more semantically aligned with their own posts than with unrelated ones, we conducted a permutation test with 1,000 random reassignments of comments while preserving the number of comments per post. A mean comparison test confirmed that comments were significantly more semantically aligned with their true posts than with randomly paired ones (p < 0.001), indicating systematic semantic alignment rather than random association. Figure~\ref{fig:comments_postsl} visualises the distribution of real and permuted similarity scores.

\subsubsection{Intent}

The intent dimension captures the communicative function of each comment. Building upon speech act theory \cite{austin1962how,searle1969speech}, we distinguished five categories: \textit{appreciation} (support or gratitude), \textit{criticism} (disagreement or dissatisfaction), \textit{inquiry} (questions or requests for information), \textit{statement} (elaboration), and \textit{forward} (amplification behaviours such as tagging other users).

A total of 508 comments were manually annotated by three annotators according to these categories (annotation guidelines in ~\ref{Annotation guide}). Inter-annotator agreement reached a Cohen's kappa $\kappa=0.63$, indicating substantial agreement. The labelled dataset was split into a development set (N = 246) used to generate synthetic comments and a held-out test set (N = 262) used to evaluate the final model.

Intent classification in Dutch social media contexts poses a low-resource challenge, as labelled datasets for fine-grained communicative intent are scarce. To address this, we adopted a data augmentation approach using large language models (LLMs), which have been shown to generate high-quality synthetic training data in low-resource settings \cite{ubani2023zeroshotdataauggeneratingaugmentingtraining}. Using the development set (N = 246) as a seed, we generated 13,000 synthetic training examples via few-shot prompting with Claude Sonnet 4.5 (see ~\ref{SDG}), preserving the developer set intent distribution.

Synthetic data was generated for all categories except \textit{Forward}. Due to its distinctive structural characteristics, this category was identified using a rule-based approach: comments beginning with a tagged user mention were classified as \textit{Forward}. As this category was structurally defined rather than semantically inferred, its classification performance reflected rule-based identification rather than model prediction. The full intent detection pipeline is visualized in \ref{Intent_detection_pipeline}

We fine-tuned three Dutch transformer models on the synthetic training data:
\begin{enumerate}
\item \texttt{GroNLP/bert-base-dutch-cased}\footnote{\url{https://huggingface.co/GroNLP/bert-base-dutch-cased}} \cite{devries2019bertje}
\item \texttt{DTAI-KULeuven/robbert-2023-dutch-large}\footnote{\url{https://huggingface.co/DTAI-KULeuven/robbert-2023-dutch-large}} \cite{delobelle-etal-2020-robbert}
\item \texttt{MoritzLaurer/mDeBERTa-v3-base-mnli-xnli}\footnote{\url{https://huggingface.co/MoritzLaurer/mDeBERTa-v3-base-mnli-xnli}} \cite{He2021DeBERTaV3ID}
\end{enumerate}

All comments were preprocessed using a lightweight normalisation pipeline that removed noise while preserving semantic cues such as casing, punctuation, and emojis; URLs, email addresses, and numbers were replaced with placeholders (\texttt{<URL>}, \texttt{<EMAIL>}, \texttt{<NUMBER>}). Each model was fine-tuned for four epochs with a learning rate of $10^{-5}$. \textit{RobBERT-2023-Dutch-Large} achieved the best accuracy on the development set of 80\% (\ref{IntentDevSetPrformance}) and was designated as the tiebreaker in the ensemble. Majority voting was applied across the three models; when all three disagreed, the tiebreaker model determined the final label. The ensemble achieved an overall accuracy of 0.85 on the held-out test set (N = 262); Table~\ref{tab:classification_performance} reports per-category classification performance. Across the full dataset (N = 3,197), the trained transformer model classified criticism as the most prevalent communicative intent (29.9\%), followed by statements (27.9\%) and inquiries (22.5\%), whereas forwarding behaviour (14.7\%) and appreciation (5.1\%) were less frequent.

\subsubsection{Sentiment}
The sentiment dimension captured the affective tone of comments, classified into \textit{positive}, \textit{neutral}, and \textit{negative} categories. Sentiment analysis is well established in opinion mining research \cite{liu2012} and is central to understanding how tenants emotionally orient themselves towards organisational communication, whether they express support, indifference, or dissatisfaction.
Given the absence of domain-specific labelled data for Dutch housing discourse, we adopted an ensemble approach, combining the predictions of the three best-performing models evaluated on a hand-labelled test dataset of 262 comments annotated by two raters (Cohen's kappa $\kappa = 0.73$). Performance metrics for all evaluated models are reported in \ref{Sentiment Scores} The three selected models are:

\begin{enumerate}
\item \texttt{nlptown/bert-base-multilingual-uncased-sentiment}\footnote{\url{https://huggingface.co/nlptown/bert-base-multilingual-uncased-sentiment}}
\item \texttt{clips/republic}\footnote{\url{https://huggingface.co/clips/republic}}
\item \texttt{citizenlab/twitter-xlm-roberta-base-sentiment-finetunned}\footnote{\url{https://huggingface.co/citizenlab/twitter-xlm-roberta-base-sentiment-finetunned}}
\end{enumerate}
Predictions were aggregated using majority voting: if at least two models agreed on a label, that label was assigned; in cases of full disagreement, the comment was labelled \textit{neutral} by default. The ensemble model achieved an overall accuracy of 0.72 on the held-out test set of 262 comments; Table~\ref{tab:classification_performance} reports classification performance across sentiment categories. In the full dataset (N = 3,197), the sentiment model classified most comments as neutral (43\%), followed by negative sentiment (39\%), while positive sentiment was least frequent (18\%). 

\begin{table}[t]
\centering
\begin{tabular}{llcccc}
\hline
\textbf{Task} & \textbf{Class} & \textbf{Precision} & \textbf{Recall} & \textbf{F1-score} & \textbf{Support} \\
\hline

\multirow{5}{*}{Intent}
 & Appreciation & 0.80 & 0.57 & 0.67 & 7 \\
 & Criticism    & 0.73 & 0.96 & 0.83 & 76 \\
 & Forward      & 1.00 & 1.00 & 1.00 & 37 \\
 & Inquiry      & 0.85 & 0.95 & 0.90 & 43 \\
 & Statement    & 0.94 & 0.69 & 0.80 & 99 \\
\cline{2-6}
 & \textbf{Accuracy} & \multicolumn{4}{c}{0.85 (N = 262)} \\

\hline

\multirow{3}{*}{Sentiment}
 & Negative & 0.75 & 0.74 & 0.74 & 98 \\
 & Neutral  & 0.76 & 0.66 & 0.70 & 114 \\
 & Positive & 0.57 & 0.85 & 0.68 & 50 \\
\cline{2-6}
 & \textbf{Accuracy} & \multicolumn{4}{c}{0.72 (N = 262)} \\

\hline
\end{tabular}
\caption{Performance of the majority-voting ensemble on intent and sentiment classification tasks, evaluated on the held-out test set.}
\label{tab:classification_performance}
\end{table}

\subsubsection{Clustering: Intent X Sentiment X Relatedness}

Unlike prior work that relied on single-dimensional representations, our approach combines probabilistic intent distributions, ordinal sentiment, and continuous semantic similarity into a unified feature space, enabling the identification of multidimensional discourse configurations.

With the three comment-level dimensions estimated, we derived the discourse typology using unsupervised clustering. Each comment was represented by seven features: five intent probabilities, a sentiment score, and a relatedness score. Intent was operationalised as the average predicted probability across three transformer models for four categories (\textit{Appreciation}, \textit{Criticism}, \textit{Inquiry}, and \textit{Statement}). The \textit{Forward} category was identified using a rule-based binary indicator (1 if present, 0 otherwise). Sentiment was included as a normalised continuous score due to differences in sentiment model output schemes, while semantic relatedness was included as a continuous score.

All features were standardised to a zero mean and unit variance before clustering to ensure comparability across dimensions. We applied k-means clustering and a fixed random seed (\texttt{random\_state=42}) to ensure reproducibility. Alternative methods were evaluated: hierarchical agglomerative clustering yielded highly imbalanced solutions, whereas DBSCAN was sensitive to the density parameter and failed to produce stable partitions. To determine the optimal number of clusters, candidate solutions for $k = 2$ to $k = 10$ were evaluated using the silhouette score, which balances within-cluster cohesion and between-cluster separation. The score was maximised at $k = 6$ (~\ref{Clustering: Sil_Scores}), and this solution was retained as the final discourse typology. Cluster characteristics were interpreted using cluster-level mean values of all features, including intent probabilities, sentiment, and semantic relatedness.

\subsection{Communication Design and Organisational Characteristics}
Having derived the discourse typology, we examined two classes of predictors identified in Section~\ref{org_comm}: post-level communication design and organisational characteristics. Specifically, we assessed how these factors were associated with variation in discourse types using a single multinomial logistic regression model, with discourse type as the dependent variable. 

Communication design was operationalised using four post-level features. Post length was categorised as very short ($<$50 words), short (50-150 words), and medium-long ($>$150 words), serving as a proxy for informational depth. Lexical diversity, measured as the ratio of unique words to total words and binarised at the 33rd percentile into low and medium-high categories, reflected linguistic complexity. Question presence captured whether posts explicitly invited interaction, while URL presence indicated whether posts directed users to external information. Together, these features represented core dimensions of communication design: length, complexity, interactivity, and informational referencing.

Organisational characteristics were operationalised using three indicators collected from Aedes-benchmark data\footnote{Aedes-benchmark is a benchmarking system created by the Dutch housing association umbrella organisation Aedes to compare the performance of housing associations.} for the top 50 housing associations by comment count: total housing stock, average rent as a proxy for affordability, and tenant satisfaction score. The total housing stock was binarised at the 33rd percentile into low and medium-high categories. Affordability and tenant satisfaction were inverted variables; a lower raw value indicated a better outcome and was therefore binarised at the 67th percentile, such that medium and high denoted a more favourable level.

Both post-level and organisational features were derived from the same dataset of 648 unique posts and 2,882 comments across the top 50 housing associations. All features were included simultaneously in a single combined multinomial logistic regression model, with \textit{content sharing} as the reference discourse type. Table~\ref{tab:combined_features} summarises the distribution of all features across the dataset.

\begin{table}[t]
\centering
\small
\begin{tabular}{llrr}
\toprule
\multicolumn{4}{c}{\textbf{Panel A: Post-Level Features}} \\
\midrule
\textbf{Feature} & \textbf{Category} & \textbf{Count} & \textbf{\%} \\
\midrule
\multirow{3}{*}{Post Length}
  & Very short ($<$50 words)   & 213 & 32.9 \\
  & Short (50--150 words)      & 389 & 60.0 \\
  & Medium-Long ($>$150 words) & 46  & 7.1  \\
\addlinespace
\multirow{2}{*}{Lexical Diversity}
  & Low ($\leq$33rd percentile)         & 191 & 29.5 \\
  & Medium-High ($>$33rd percentile)   & 457 & 70.5 \\
\addlinespace
\multirow{2}{*}{Question Presence}
  & No  & 358 & 55.2 \\
  & Yes & 290 & 44.8 \\
\addlinespace
\multirow{2}{*}{URL Presence}
  & No  & 401 & 61.9 \\
  & Yes & 247 & 38.1 \\
\midrule
\multicolumn{4}{c}{\textbf{Panel B: Organisational Characteristics}} \\
\midrule
\textbf{Feature} & \textbf{Category} & \textbf{Count} & \textbf{\%} \\
\midrule
\multirow{2}{*}{Total Housing Stock}
  & Low ($\leq$33rd percentile)              & 295 & 45.5 \\
  &Medium-High  ($>$33rd percentile)      & 353 & 54.5 \\
\addlinespace
\multirow{2}{*}{Affordability Level}
  & Low   ($>$67th percentile)            & 223 & 34.4 \\
  &Medium-High ($\leq$67th percentile)    & 425 & 65.6 \\
\addlinespace
\multirow{2}{*}{Tenant Satisfaction}
  & Low   ($>$67th percentile           & 170 & 26.2 \\
  &Medium-High ($\leq$67th percentile)      & 478 & 73.8 \\
\bottomrule
\end{tabular}
\caption{Summary of post-level features and organisational characteristics used in the multinomial logistic regression. The top 50 organisations by comment count were considered, resulting in a dataset of 648 unique posts and 2,882 comments.}
\label{tab:combined_features}
\end{table}

\section{Results and Discussion}
\label{results}

\subsection{Discourse Typology}

\begin{figure}[t]
\centering
\includegraphics[scale = 0.22]{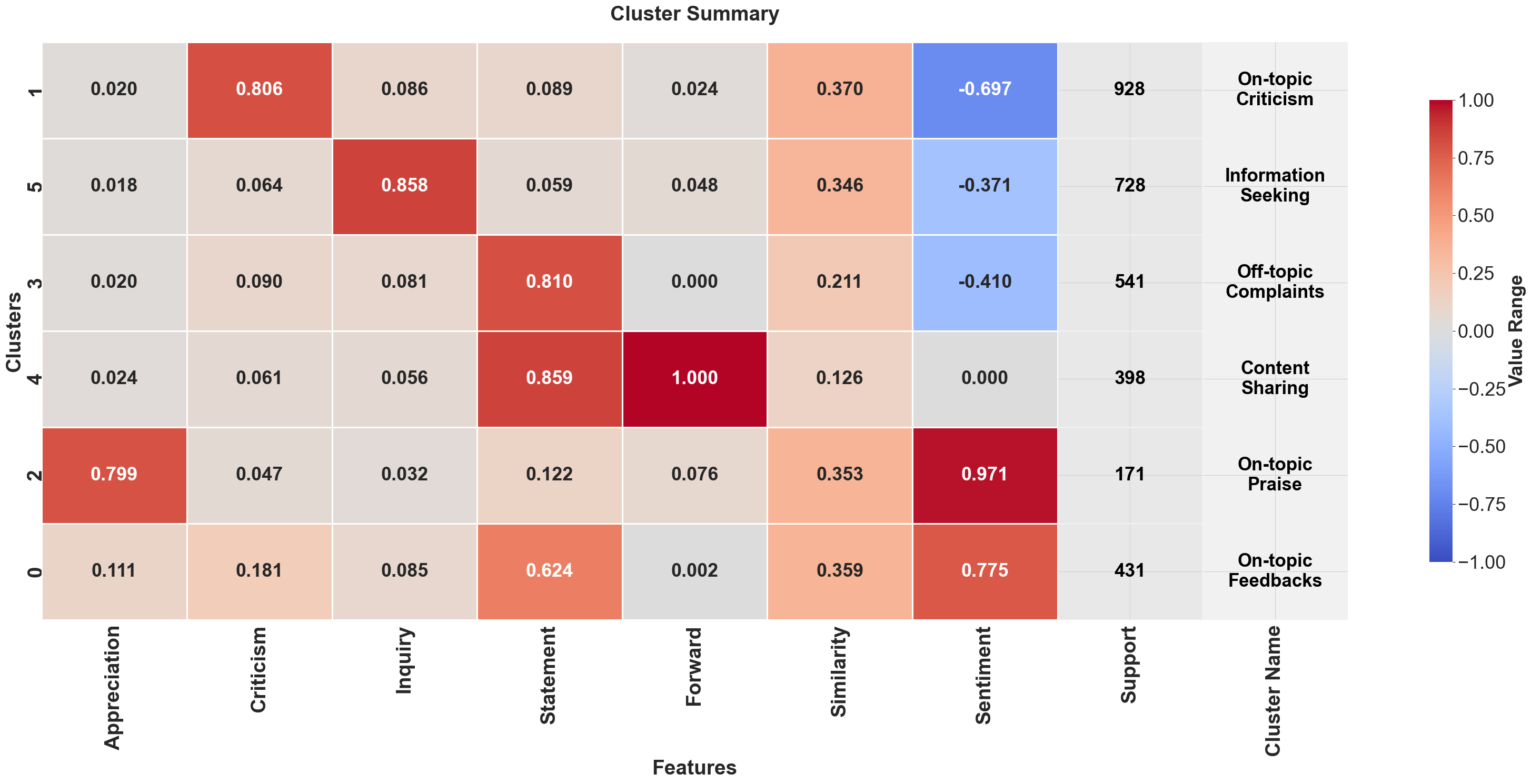}
\caption{Cluster-level summary of intent probabilities, similarity, and sentiment.}
\label{fig:cluster_heatmap}
\end{figure}

Figure~\ref{fig:cluster_heatmap} presents the cluster-level distribution of semantic relatedness, sentiment, and communicative intent. The six-cluster solution revealed differentiated and internally consistent configurations, indicating that citizen discourse was structured along multiple dimensions rather than forming a homogeneous distribution.

The identified discourse types -- on-topic feedback, on-topic criticism, on-topic praise, off-topic complaints, content sharing, and information seeking -- correspond to distinct combinations of intent, affect, and topical alignment. As shown in Figure~\ref{fig:cluster_heatmap}, on-topic clusters exhibited high semantic relatedness but diverged in sentiment and communicative function, separating evaluative (criticism and praise) from interactive (feedback and inquiry) engagement. Off-topic complaints formed a distinct configuration characterised by low relatedness and strongly negative, expressive intent, while content sharing reflected low-intensity, redistribution-oriented behaviour.

These results demonstrate that discourse types emerge as structured configurations of the three dimensions, spanning a continuum from substantive, issue-focused engagement to passive or tangential participation. The presence of a stable off-topic cluster further indicates that expressive, weakly related contributions constitute a recognisable mode of engagement rather than residual noise. More broadly, these findings validate the multidimensional operation of discourse: the identified clusters arise directly from the joint distribution of semantic relatedness, sentiment, and communicative intent, confirming that these dimensions capture meaningful and recurrent patterns of citizen engagement.

\begin{figure}[t]
\centering
\includegraphics[width=0.99\linewidth]{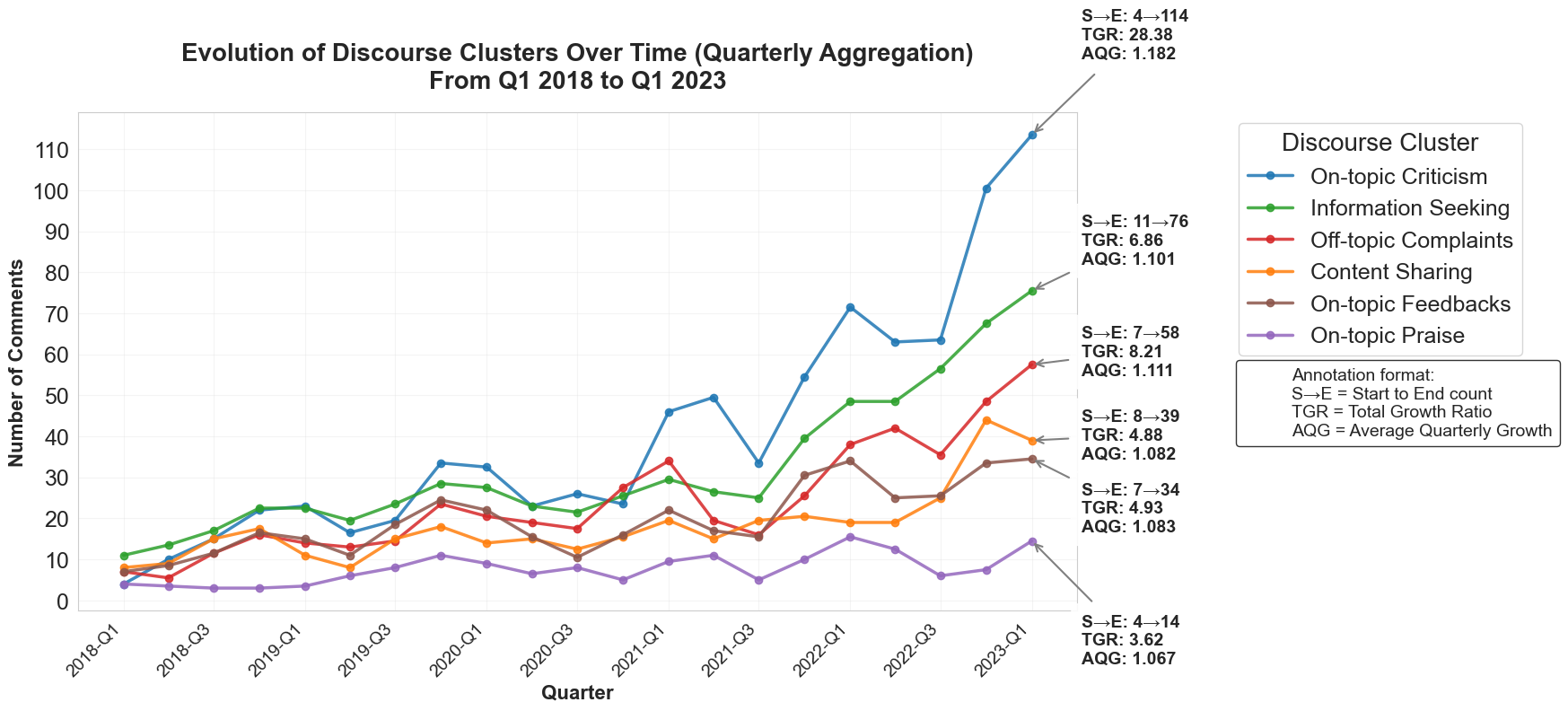}
\caption{Evolution of the discourse by type over time (quarterly aggregation from Q1 2018 to Q1 2023). On-topic Criticism and Information Seeking show the most pronounced growth, particularly from 2020 onwards, while other discourse types remain relatively stable.}
\label{fig:dis_evol}
\end{figure}

Temporal dynamics, shown in Figure~\ref{fig:dis_evol}, reinforce the above interpretation. The share of on-topic criticism increased markedly from 2020 onwards, alongside a rise in information-seeking, while content sharing and on-topic praise remained comparatively stable. This shift reflects a redistribution of engagement towards more evaluative and inquiry-oriented configurations rather than a uniform increase in activity. Although causal attribution is not possible, the timing coincides with the COVID-19 period and increased reliance on digital communication, alongside rising issue salience in domains such as energy and sustainability.

\subsection{Communication Design and Organisational Features across Discourse Types}

\begin{figure}[t]
    \centering
    \includegraphics[width=0.7\linewidth]{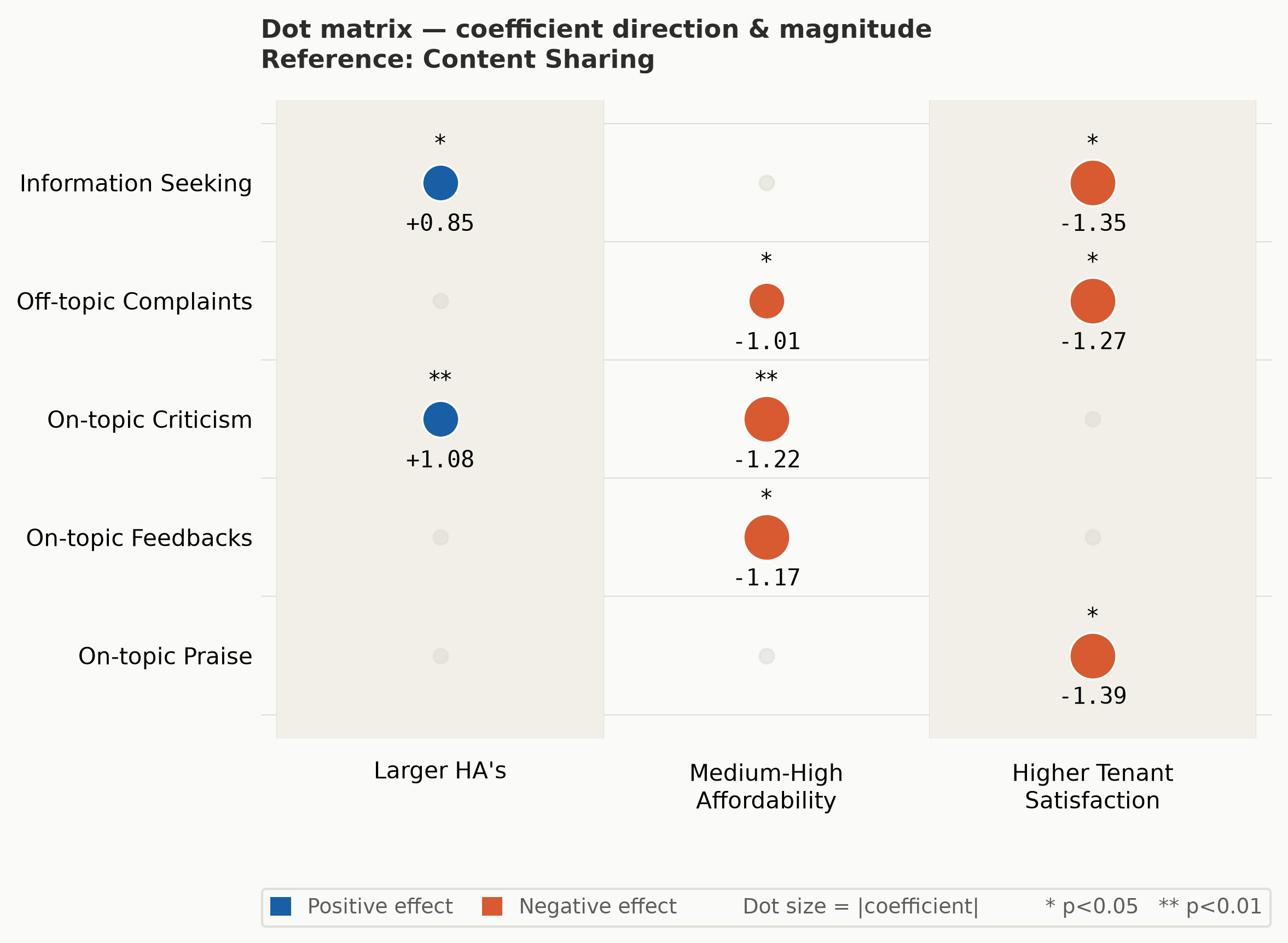}
    \caption{Statistically significant predictors of discourse type from the combined 
    multinomial logistic regression model. Only statistically significant 
    effects ($p < 0.05$) are shown. Full results are reported 
    in Table~\ref{tab:combined_mnlogit_results} in the Appendix.}
    \label{fig:mlt_results}
\end{figure}

Table~\ref{tab:combined_mnlogit_results} (see \ref{mlt_results}) presents the full results of the combined multinomial logistic regression model, with content sharing as the reference discourse type. Figure~\ref{fig:mlt_results} visualises the statistically significant effects. Overall, the model revealed that organisational characteristics systematically dominated the prediction of discourse type composition, while post-level communication design features did not independently explain meaningful variation once organisational context was controlled for.

Among organisational characteristics, affordability level was the most consistent predictor, as indicated by the prevalence of negative effects for lower-rent housing associations in Figure~\ref{fig:mlt_results}. Housing associations with medium-high affordability (i.e., lower rents) were significantly less likely to elicit on-topic feedback ($\beta = -1.17$, $p < 0.05$), on-topic criticism ($\beta = -1.22$, $p < 0.01$), and off-topic complaints ($\beta = -1.01$, $p < 0.05$) relative to content sharing, suggesting reduced evaluative and complaint-driven engagement among tenants of more affordable providers. By contrast, organisational scale showed a positive association: larger housing associations were more likely to receive on-topic criticism ($\beta = 1.08$, $p < 0.01$) and information-seeking comments ($\beta = 0.85$, $p < 0.05$), indicating that greater scale is associated with more substantive engagement. Finally, tenant satisfaction was negatively associated with on-topic praise ($\beta = -1.39$, $p < 0.05$), off-topic complaints ($\beta = -1.27$, $p < 0.05$), and information seeking ($\beta = -1.35$, $p < 0.05$), suggesting that higher satisfaction reduces both expressive and inquiry-driven discourse relative to passive content sharing.


Taken together, the results suggest that discourse configurations in housing association communication are shaped primarily by structural organisational conditions rather than post-communication features. Affordability, organisational scale, and tenant satisfaction systematically influence how tenants engage across dimensions of sentiment, intent, and topical alignment. These findings align with earlier literature suggesting that communication design and organisational context may shape online engagement while indicating a comparatively stronger role for organisational characteristics in explaining discourse variation in this setting.

\section{Conclusion}
\label{Conclusion}
This study proposed a data-driven framework to analyse the rich information embedded in discourse on organizational social media, motivated by the need to move beyond aggregate engagement metrics toward a more fine-grained understanding of citizen responses. We applied the framework to analyse tenant responses to Dutch housing association sustainability communication on Facebook.

Three key conclusions emerge from this analysis. First, social media engagement is not uniform but exhibits a clear structure. Tenant responses cluster into six distinct discourse types: on-topic feedback, on-topic criticism, on-topic praise, off-topic complaints, content sharing, and information seeking. Together, these discourse types reflect affective, behavioural, and topical dimensions of engagement, while remaining largely aligned with organisational communication. This finding suggests that organisations retain a degree of agenda-setting capacity even in open digital environments, as they continue to influence the focus of online discourse.

Second, communication design does not independently explain meaningful variation in discourse type composition once organisational characteristics are accounted for. The majority of posts are short to medium in length, with over 90\% falling below 150 words, and nearly half contain explicit questions, reflecting a broadly dialogic character across the dataset. Yet these structural properties do not translate into systematic differences in discourse type composition, suggesting that how a post is written matters less than the organisational context in which it appears. Third, and most substantially, organisational characteristics systematically shape discourse patterns. Housing providers that are larger in scale tend to attract more substantive and inquiry-driven engagement, while those with lower rent levels are less likely to elicit evaluative or complaint-driven responses. Higher tenant satisfaction is associated with a relative suppression of expressive and inquiry-driven discourse, with tenants appearing to default toward passive content sharing when satisfaction levels are favourable.

More broadly, this study challenges the widespread reliance on engagement metrics as proxies for public response quality. By showing that discourse is structured, multidimensional, and context-dependent, it provides a foundation for more fine-grained analysis of citizen engagement in digital governance environments. From a practical perspective, the findings suggest that communication outcomes are deeply embedded in organisational context: the same message may generate fundamentally different discourse depending on organisational scale, resource allocation, and tenant satisfaction. Effective communication strategies must therefore account for the structural position of the organisation rather than focusing solely on message design. Despite being developed in the context of Dutch public housing, the framework is readily transferable to other organisational and policy settings, offering a scalable approach to analysing how communication shapes public discourse across contexts.

Several limitations should be acknowledged. The analysis is observational and therefore does not permit causal claims regarding the relationship between communication features and discourse outcomes. In addition, the study focuses on a single platform and policy domain, which may limit the generalisability of the results. The identification of discourse types, while grounded in data-driven clustering, also involves interpretative choices, and the binarisation of organisational characteristics may conceal more nuanced relationships. These limitations point to several avenues for future research, including experimental designs that enable causal inference, applications across different platforms and policy settings, qualitative validation of discourse categories, and the use of continuous or non-linear measures of organisational characteristics. An additional extension would be to classify organisational posts by communicative function, allowing researchers to examine how communication strategies interact with organisational context in shaping public discourse.

\section{Code Availability}

The code is available \href{https://github.com/Shray15/engd_WP1_WoonPraat}{here}.

\section{Funding}
We received support from the Dutch Science Foundation (NWO) grant 403.19.230 and the Dutch enterprise agency (RVO).

\bibliography{sample}

\section{Acknowledgements}

We acknowledge support from the Dutch Science Foundation (NWO) grant 403.19.230 and the Dutch enterprise agency (RVO). We thank the employees of Woonbedrijf, especially Monique van Lent-Deelstra, for sharing their expertise, useful comments, and discussions. During the preparation of this work we used ChatGPT to assist with English language editing. After using this tool, we reviewed and edited the content as needed and take full responsibility for the content of the publication.

\section{Author contributions statement}
All authors developed the study concept and designed the investigation. SJ and IVO collected and verified the data. SJ did the data analysis and wrote the first draft of the manuscript. All authors contributed to data interpretation and reviewed and edited the manuscript. All authors consented to submit the manuscript. 

\section{Declaration of interests}
We declare no competing interests.

\clearpage
\appendix

\appendix

\renewcommand{\thesection}{Appendix \Alph{section}.}
\renewcommand{\thesubsection}{Appendix \Alph{section}.\arabic{subsection}}

\counterwithin{table}{section}
\counterwithin{figure}{section}
\renewcommand{\thetable}{\Alph{section}\arabic{table}}
\renewcommand{\thefigure}{\Alph{section}\arabic{figure}}

\newpage

\noindent {\large \textbf{APPENDICES}}

\section{Annotation instructions} \label{Annotation guide}

\vspace{1em}
\textbf{General Instructions for Intent Annotation} \\

\textbf{Task Objective:} \\
Assign one of the predefined intent labels to each comment based on the user's underlying purpose or emotional expression.

\vspace{0.5em}
\textbf{Possible Intent Labels:}
\begin{itemize}
  \item Criticism
  \item Appreciation
  \item Statement – Positive
  \item Statement – Negative
  \item Statement – Neutral
  \item Inquiry
  \item Forward
\end{itemize}

\textbf{How to Annotate:}
\begin{itemize}
  \item Read each comment carefully.
  \item Consider the tone, structure, and word choice to identify the intent.
  \item Select the single intent that best matches the comment.
\end{itemize}

\textbf{If unsure:}
\begin{itemize}
  \item Choose the label that reflects the dominant tone or purpose.
  \item Use the examples and patterns below to guide your decision.
  \item If the intent feels ambiguous, assign a secondary intent.
\end{itemize}

\vspace{0.5em}
\textbf{Data Structure:} \\
\textbf{Text:} Comment to be annotated \\
\textbf{Primary Intent:} Dominant intent, clear intent of the comment \\
\textbf{Secondary Intent:} If the comment feels ambiguous, assign a secondary intent along with the primary intent.

\vspace{1em}
\textbf{1. Criticism} \\
\textbf{Definition:} The author expresses dissatisfaction, disapproval, or negative judgment about a situation, person, organization, or decision. \\
\textit{Note: The comment should blame someone (criticize a particular subject/entity) for their actions or behavior.}

\textbf{Common Patterns:}
\begin{itemize}
  \item Use of negative words
  \item Tone of frustration or disappointment
  \item Blame or call-outs – must be specific
\end{itemize}

\textbf{Trigger Phrases \& Words:}
\begin{itemize}
  \item This is unacceptable…
  \item It's a shame that…
  \item Why would they…?
  \item No one is doing anything about…
\end{itemize}

\vspace{0.5em}
\textbf{2. Appreciation} \\
\textbf{Definition:} The author expresses approval, gratitude, or positive feedback towards a person, service, or decision. There should be an entity that is being praised.

\textbf{Common Patterns:}
\begin{itemize}
  \item Positive tone and sentiment
  \item Use of words like “great,” “thank you,” “happy,” “nice”
\end{itemize}

\textbf{Trigger Phrases \& Words:}
\begin{itemize}
  \item Thank you for…
  \item Great work by…
  \item Well done…
\end{itemize}

\vspace{0.5em}
\textbf{3. Statement – Positive} \\
\textbf{Definition:} The author shares an observation, fact, or opinion with a positive tone, but does not praise any entity.

\textbf{Common Patterns:}
\begin{itemize}
  \item Informative or descriptive tone
  \item Often factual or narrative
\end{itemize}

\textbf{Trigger Phrases \& Words:}
\begin{itemize}
  \item Beautiful flowers…
  \item This works well…
  \item Exactly what I needed.
\end{itemize}

\vspace{0.5em}
\textbf{4. Statement – Negative} \\
\textbf{Definition:} The author shares an observation, fact, or opinion with a negative tone, but does not blame anyone.

\textbf{Trigger Phrases \& Words:}
\begin{itemize}
  \item This does not work…
  \item This is not what I expected.
  \item There's a serious issue with…
  \item Clearly something's wrong…
\end{itemize}

\vspace{0.5em}
\textbf{5. Statement – Neutral} \\
\textbf{Definition:} The author shares an observation, fact, or opinion in a neutral tone, without clear emotional bias.

\textbf{Trigger Phrases \& Words:}
\begin{itemize}
  \item It appears that…
  \item This is what happened…
  \item I noticed that…
  \item As far as I can tell…
\end{itemize}

\vspace{0.5em}
\textbf{6. Inquiry} \\
\textbf{Definition:} Includes both:
\begin{itemize}
  \item \textbf{Questions:} Asking for information, clarification, or solution
  \item \textbf{Requests:} Asking for action or help
\end{itemize}

\textbf{Common Patterns:}
\begin{itemize}
  \item Question format (with or without a question mark)
  \item Tone of curiosity or confusion
  \item Implicit questions (indirect inquiries)
\end{itemize}

\textbf{Trigger Phrases \& Words:}
\begin{itemize}
  \item Why is…?
  \item Can someone tell me…
  \item Is this still available?
  \item What happened to…?
  \item I wonder…
  \item Maybe you could…
  \item It would be nice to…
\end{itemize}

\vspace{0.5em}
\textbf{7. Forward} \\
\textbf{Definition:} The author tags another person or shares the post to inform someone else. Sentiment is mostly neutral or positive.

\textbf{Trigger:} Presence of a \texttt{<PERSON>} tag in the beginning of the comment implies Forward intent.

\newpage
\section{Synthetic Dataset generation}
\label{SDG}

A few-shot prompting technique was used to generate a Synthetic Dataset using Claude Sonnet 4.5 to train the BERT models for intent recognition. Below is an example of the prompt used to generate Synthetic comments for the intent \textit{Criticism}.

\begin{tcolorbox}[title={Few-shot Prompt to generate Synthetic Dataset}]
You are a synthetic dataset generator. Generate 200 unique comments for the intent "Criticism" in the Dutch language. The comments should represent human language and mimic as if the user is posting on Facebook housing association posts. \\[1ex]

Consider these examples for reference: \\
1. Klopt toch niks van, na 3 jaar een woning? Een ander reageert zich suf en krijgt helemaal niks, na 8 jaar ingeschreven te staan en dus ook al 8 jaar reageert op lotingen. \\
2. Op een woning met achterstallig onderhoud. En daarna flink betalen aan de vve. Dat geintje hebben ze ook bij Lauwers gedaan. Resultaat. Alles mooi opgeknapt behalve de verkochte flat. \\
3. Mijn huis zit vol schimmel me kleding schoenen alles in huis gaat er aan we zijn ook altijd ziek 
\end{tcolorbox}

Similarly, comments were generated for other intents, except \textit{Forward}, by replacing the intent in the prompt, along with examples for reference (no posts were inserted in the prompt). Comments were generated iteratively.

The synthetic comments were generated in an iterative process, where models were trained and evaluated on the developer set after each loop. In every iteration, additional comments were produced by leveraging new examples from the developer set, thereby gradually expanding the synthetic dataset. This approach was intended to improve the diversity of the dataset, enabling the models to better capture the nuances present in the comments.
 
\newpage
\section{Intent Detection Pipeline} 
\label{Intent_detection_pipeline}
\begin{figure}[ht]
    \centering
    \includegraphics[width=0.9\linewidth]{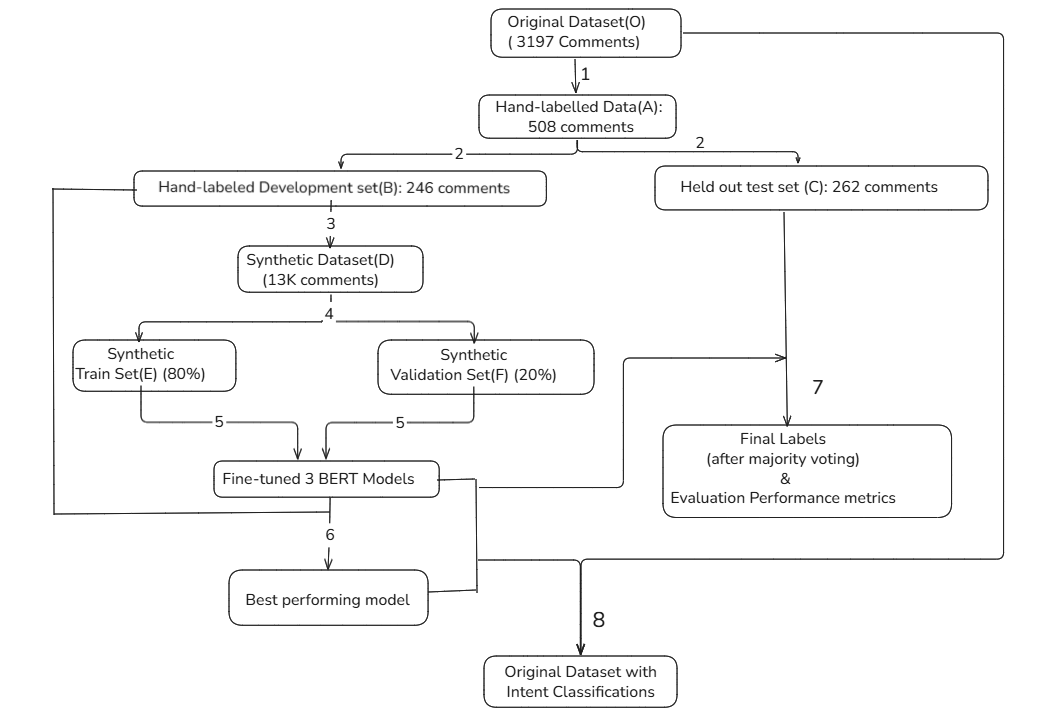}
    \caption{Overview of the intent detection pipeline.}
    \label{app:intent_pipeline}
\end{figure}

The full intent detection pipeline diagram is visualised in Figure~\ref{app:intent_pipeline}.

\newpage
\section{Intent Classification performance of individual models on the development dataset.}\label{IntentDevSetPrformance}

\begin{table}[h]
\centering
\begin{tabular}{lcccccccccccc}
\hline
\textbf{Class} & \multicolumn{3}{c}{\textbf{BERTje}} & \multicolumn{3}{c}{\textbf{DeBERTaV3}} & \multicolumn{3}{c}{\textbf{Robbert-2023}} \\
               & Precision & Recall & F1-Score & Precision & Recall & F1-Score & Precision & Recall & F1-Score \\
\hline
Appreciation  & 0.65   &   1.00   &   0.79     &  0.61   &   0.73   &  0.67    & 0.78    &  0.47   &   0.58    \\
Criticism     & 0.70   &   0.73    &  0.71     &  0.64   &   0.89  &   0.75      & 0.65   &   0.93  &    0.76      \\
Inquiry       & 0.63    &  0.89    &  0.73     & 0.76    &  0.85   &   0.80     & 0.82    &  0.87    &  0.84     \\
Statement     &  0.83    &  0.52  &    0.64     & 0.89   &   0.58  &    0.70      &0.86    &  0.65   &   0.74  \\
\hline
\textbf{Accuracy} & \multicolumn{3}{c}{0.74} & \multicolumn{3}{c}{0.78} & \multicolumn{3}{c}{0.80} \\
\hline
\end{tabular}
\caption{Individual Performance Comparison of Intent Classification Models on dev set.}
\label{tab:model_comparison_dev_set}
\end{table}

\section{Evaluation of Sentiment analysis models on Hand-labeled sentiment data} \label{Sentiment Scores}

Table.\href{tab:ind_sentiment-scores}{\ref{tab:ind_sentiment-scores}} depicts the accuracy of the individual sentiment analysis models on the hand-labeled sentiment dataset($A$)

\begin{table}[h!]
\centering
\begin{tabular}{|p{8.5cm}|c|}
\hline
\textbf{Model} & \textbf{Score} \\
\hline
DTAI-KULeuven/robbert-v2-dutch-sentiment & 0.42 \\
pdelobelle/robbert-v2-dutch-base & 0.34 \\
DTAI-KULeuven/robbertje-merged-dutch-sentiment & 0.39 \\
nlptown/bert-base-multilingual-uncased-sentiment & \textbf{0.62} \\
GroNLP/bert-base-dutch-cased & 0.42 \\
citizenlab/twitter-xlm-roberta-base-sentiment-finetunned & \textbf{0.64} \\
BramVanroy/xlm-roberta-base-hebban-reviews & 0.53 \\
BramVanroy/bert-base-multilingual-cased-hebban-reviews & 0.37 \\
BramVanroy/robbert-v2-dutch-base-hebban-reviews & 0.39 \\
clips/republic & \textbf{0.70} \\
\hline
\end{tabular}
\caption{Performance scores of various sentiment models}
\label{tab:ind_sentiment-scores}
\end{table}

Here, the annotator is asked to assign a comment to three Statement types for Sentiment detetction purpose. However, during intent detection, these three Statement splits are unified.



\newpage
\section{Clustering: Silhouette Scores and Graph.} \label{Clustering: Sil_Scores}

\begin{figure}[ht] 
\centering 
\includegraphics[width=0.8\linewidth]{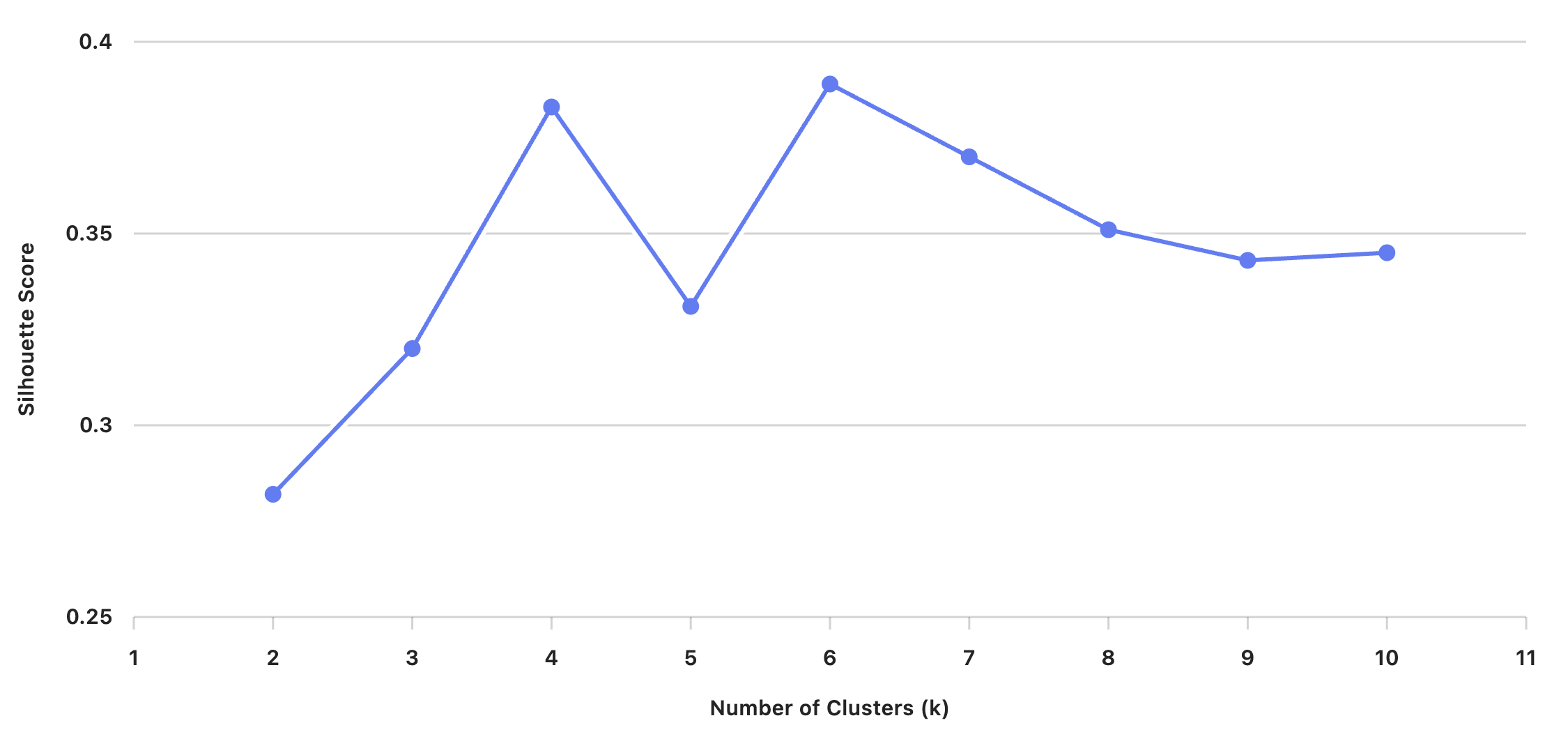} 
\caption{Silhouette scores for \(k = 2\) to \(k = 10\). The highest score is achieved at \(k = 6\), indicating that this is the optimal number of clusters for the discourse classification.} 
\label{fig:silhouette_scores} 
\end{figure}

The silhouette score for clustering is visualized in the figure. \ref{fig:silhouette_scores}. The highest score is achieved at \(k = 6\), indicating that this is the optimal number of clusters for the discourse classification.

\newpage
\section{Multinomial Logistic Regression results of combining organizational and communication design features} \label{mlt_results}

\begin{table}[ht]
\centering
\small
\begin{tabular}{lccccc}
\toprule
& \multicolumn{5}{c}{\textbf{Discourse Type (Reference: Content Sharing, Cluster 4)}} \\
\cmidrule(lr){2-6}
\textbf{Feature} & \textbf{On-topic} & \textbf{On-topic} & \textbf{On-topic} & \textbf{Off-topic} & \textbf{Information} \\
& \textbf{Feedbacks} & \textbf{Criticism} & \textbf{Praise} & \textbf{Complaints} & \textbf{Seeking} \\
& (Cluster 0) & (Cluster 1) & (Cluster 2) & (Cluster 3) & (Cluster 5) \\
\midrule
\textbf{Intercept}
& \textbf{2.007*} & \textbf{2.226*} & 1.431 & 1.757 & \textbf{2.293**} \\
& (0.931) & (0.868) & (1.128) & (0.949) & (0.877) \\
\addlinespace
\multicolumn{6}{l}{\textit{Lexical Diversity (Ref: Low)}} \\
Medium--High
& $-0.095$ & $-0.038$ & $-1.007$ & $-0.338$ & $0.050$ \\
& (0.555) & (0.514) & (0.650) & (0.567) & (0.522) \\
\addlinespace
\multicolumn{6}{l}{\textit{Post Length (Ref: Very Short)}} \\
Short
& $0.556$ & $0.556$ & $0.627$ & $0.331$ & $0.422$ \\
& (0.456) & (0.413) & (0.633) & (0.472) & (0.417) \\
Medium--Long
& $0.199$ & $0.350$ & $-0.360$ & $-0.061$ & $-0.105$ \\
& (0.946) & (0.865) & (1.211) & (0.972) & (0.896) \\
\addlinespace
\multicolumn{6}{l}{\textit{Post-Level Features}} \\
Question Presence
& $0.338$ & $0.614$ & $-0.340$ & $0.756$ & $0.568$ \\
& (0.428) & (0.390) & (0.564) & (0.435) & (0.396) \\
URL Presence
& $-0.560$ & $0.014$ & $0.325$ & $0.147$ & $0.125$ \\
& (0.422) & (0.371) & (0.522) & (0.420) & (0.377) \\
\addlinespace
\multicolumn{6}{l}{\textit{Total Housing Stock (Ref: Low)}} \\
Medium--High
& $0.750$ & \textbf{1.080**} & $0.652$ & $0.794$ & \textbf{0.853*} \\
& (0.430) & (0.395) & (0.532) & (0.440) & (0.401) \\
\addlinespace
\multicolumn{6}{l}{\textit{Affordability Level (Ref: Low)}} \\
Medium--High
& \textbf{-1.173*} & \textbf{-1.223**} & $-0.824$ & \textbf{-1.009*} & $-0.821$ \\
& (0.490) & (0.455) & (0.618) & (0.506) & (0.466) \\
\addlinespace
\multicolumn{6}{l}{\textit{Tenant Satisfaction Level (Ref: Low)}} \\
Medium--High
& $-1.042$ & $-0.792$ & \textbf{-1.392*} & \textbf{-1.266*} & \textbf{-1.354*} \\
& (0.560) & (0.529) & (0.652) & (0.567) & (0.526) \\
\midrule
\multicolumn{6}{l}{\textbf{Model Statistics}} \\
\multicolumn{6}{l}{$N = 648$ \quad Pseudo $R^2 = 0.044$ \quad} \\
\multicolumn{6}{l}{Method: Maximum Likelihood Estimation} \\
\bottomrule
\end{tabular}
\caption{Multinomial logistic regression results: combined post-level and organisational features predicting discourse type. Coefficients represent log odds relative to Content Sharing (Cluster 4).
$^{.}p<0.10$, $^{*}p<0.05$, $^{**}p<0.01$, $^{***}p<0.001$.}
\label{tab:combined_mnlogit_results}
\end{table}

\clearpage
\end{document}